\documentclass[aps,prb,10pt,twocolumn,superscriptaddress]{revtex4-1}
\usepackage{graphicx}
\usepackage{amsmath}
\usepackage[squaren]{SIunits}
\usepackage{amssymb}
\usepackage{amsfonts}
\usepackage[english]{babel}
\usepackage{fancyhdr}
\usepackage{enumerate}
\begin{document}

\title{Structure-dependent optical and electrical transport properties of nanostructured Al-doped ZnO}
\author{P. Gondoni}
\affiliation{Dipartimento di Energia and NEMAS $–$ Center for NanoEngineered Materials and Surfaces, Politecnico di Milano, via Ponzio 34/3, 20133 Milano (Italy)}
\author{M. Ghidelli}
\affiliation{Dipartimento di Energia and NEMAS $–$ Center for NanoEngineered Materials and Surfaces, Politecnico di Milano, via Ponzio 34/3, 20133 Milano (Italy)}
\author{F. Di Fonzo}
\affiliation{Center for Nano Science and Technology @Polimi, Istituto Italiano di Tecnologia, Via Pascoli 70/3, 20133 Milano (Italy)}
\author{M. Carminati}
\affiliation{Dipartimento di Elettronica e Informazione, Politecnico di Milano, Piazza Leonardo da Vinci 32, 20133 Milano (Italy)}
\author{V. Russo}
\affiliation{Dipartimento di Energia and NEMAS $–$ Center for NanoEngineered Materials and Surfaces, Politecnico di Milano, via Ponzio 34/3, 20133 Milano (Italy)}
\author{A. Li Bassi}
\affiliation{Dipartimento di Energia and NEMAS $–$ Center for NanoEngineered Materials and Surfaces, Politecnico di Milano, via Ponzio 34/3, 20133 Milano (Italy)}
\affiliation{Center for Nano Science and Technology @Polimi, Istituto Italiano di Tecnologia, Via Pascoli 70/3, 20133 Milano (Italy)}
\author{C.S. Casari}
\affiliation{Dipartimento di Energia and NEMAS $–$ Center for NanoEngineered Materials and Surfaces, Politecnico di Milano, via Ponzio 34/3, 20133 Milano (Italy)}
\affiliation{Center for Nano Science and Technology @Polimi, Istituto Italiano di Tecnologia, Via Pascoli 70/3, 20133 Milano (Italy)}

\begin{abstract}
The structure-property relation of nanostructured Al-doped ZnO
thin films has been investigated in detail through a systematic
variation of structure and morphology, with particular emphasis on
how they affect optical and electrical properties. A variety of
structures, ranging from compact polycristalline films to
mesoporous, hierarchically organized cluster assemblies, are grown
by Pulsed Laser Deposition at room temperature at different oxygen
pressures. We investigate the dependence of functional properties
on structure and morphology and show how the correlation between
electrical and optical properties can be studied to evaluate
energy gap, conduction band effective mass and transport
mechanisms. Understanding these properties opens the way for
specific applications in photovoltaic devices, where optimized
combinations of conductivity, transparency and light scattering
are required.
\end{abstract}
\maketitle

\section{Introduction}
The importance of Transparent Conducting Oxides (TCOs) in several
fields, from optoelectronics to energy harvesting, has become
unquestionable over the past decades \cite{Ginley2000,
Exarhos2007, Minami2005}. More recently, economical and
environmental reasons have raised the need to move away from
indium-based TCOs \cite{Minami2008, Minami2008a, Park2009AGA} and
focus on cheaper and easily achievable solutions such as ZnO-based
materials \cite{Ozgur2005} whose interest is increased by the
possibility of employment also as nano- and mesoporous photoanodes
in dye-sensitized solar cells (DSSCs)
\cite{Martinson2009a,Zhang2009} as well as conventional TCOs in
organic and hybrid solar cells \cite{Zhang2010}. For this kind of
applications, it is a matter of primary importance to understand
the relation between structure and properties. For example, the
application in DSSCs requires porous structures with high surface
area, which are expected to have very different properties with
respect to conventional compact films; the development of TiO$_2$
photoanodes with hierarchical structure has been studied trying to
improve device performances \cite{Sauvage2010, Zhang2011}. \\
While a considerable amount of research has been devoted to optimizing the individual functional properties which determine the performance as TCOs (i.e. resistivity and optical transparency) \cite{Birkholz2003,Dong2007,Oda2010,Makino2002}, considerably less work has been done on systematically studying the structure/property relation in nanostructured systems, for doped\cite{Nasr2010} and undoped \cite{A2008} ZnO. One powerful means of obtaining a large variety of structures is constituted by Pulsed Laser Deposition (PLD), whose versatility in the growth of metal oxide nanostructures has been demonstrated \cite{Dellasega2008,Bailini2007,Fonzo2009} and whose applicability to the synthesis of Al-doped ZnO (AZO) films is well known \cite{Singh2001,Kim2002,Agura2003,Lackner2005}.
Also, although some authors have already performed systematic studies on some fundamental physical properties of AZO grown by PLD \cite{Dong2008}, no such work has been done (to the best of our knowledge) on nanostructured, porous films, nor on films grown at room temperature, whose potential compatibility with polymeric substrates is interesting for novel applications.\\
For the reasons described above, this work is focused on the properties of AZO films grown by PLD at room temperature, with particular emphasis on the correlation between structural, electrical and optical properties. \\
AZO films with a variety of structures spanning from compact to
mesoporous have been synthesized by tuning the plasma expansion
through O$_2$ background gas pressure and characterized in terms
of structure, morphology, electrical and optical properties.
Compact films exhibit low resistivity ($\rho \approx 4 \cdot
10^{-4} \Omega\,$cm) and high transparency (mean visible
transmittance 85\%) and porous films have enhanced light
scattering properties. The role played by stoichiometry and
defects (i.e. oxygen vacancies) is understood by performing
annealing treatments in air. A combined study of the optical
absorption profiles and Hall effect measurements allows us to
estimate the conduction band electron concentration, the
refractive index and the electron conduction band effective mass.

\begin{figure*}
    \centering
        \includegraphics[width=0.90\textwidth]{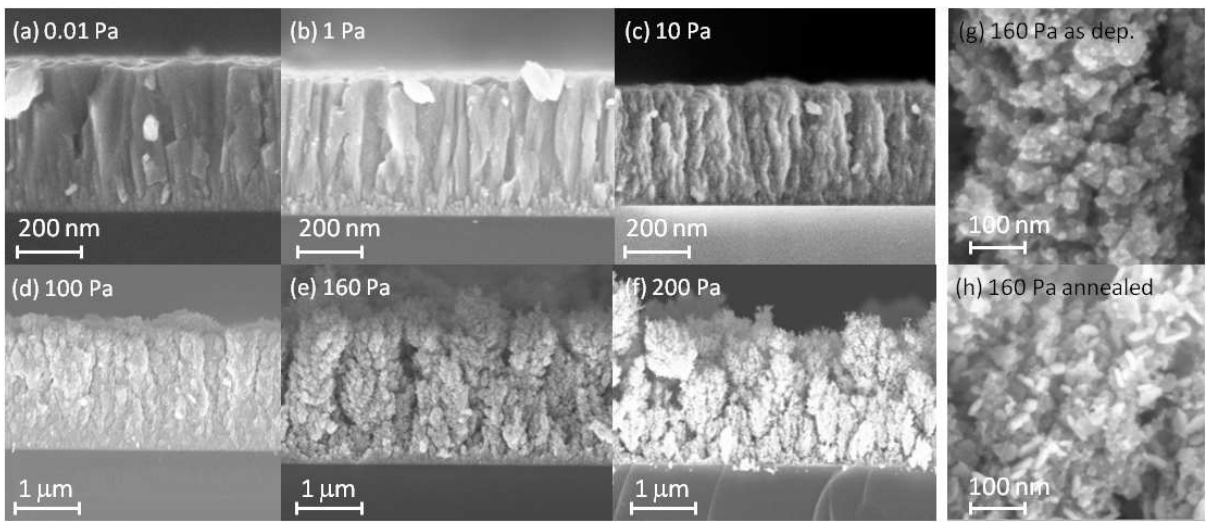}
            \caption{(a-f) Cross-sectional SEM images of AZO films grown at different O$_2$ pressures.
    (g-h) High resolution SEM images of a sample grown at 160 Pa O$_2$ before (g) and after (h) annealing in air.}
    \label{fig:sem}
\end{figure*}

\section{Experimentals}
AZO thin films were grown by PLD at room temperature, in the presence of O$_2$ atmosphere. An Al$_{2}$O$_{3}$(2\%wt.):ZnO
solid target was ablated by a ns-pulsed laser (Nd:YAG 4th harmonic,
$\lambda$=266 nm, pulse repetition rate 10 Hz, pulse duration $\sim$ 6 ns). The
fluence on the target was 1 Jcm$^{-2}$, the target-to-substrate distance was 50 mm and the background gas
(O$_{2}$) pressure was varied from 0.01 Pa to 200 Pa. The number of laser shots was 21600 for pressures up to 10 Pa, whereas for higher pressures it was adjusted in order to maintain a constant film thickness. Deposition rates were measured with a quartz microbalance. The substrates (Si(100) and soda-lime glass) were cleaned in a RF plasma of Ar ions (accelerated by a 100 V potential) prior to deposition to improve film adhesion.
Film morphology and structure were characterized by means of Scanning Electron Microscopy (Zeiss SUPRA 40 field-emission SEM, equipped with Oxford Energy Dispersive X-Ray Spectroscopy, EDXS) on samples grown on silicon. 4-point probe and Hall effect electrical measurements (Keithley 2400 Source-Measure, Agilent 34970A Voltage meter, Ecopia 0.55T magnet kit) were carried out by applying silver paint contacts on the top surface where possible (compact films), whereas the electrical properties of porous samples were probed at the substrate/film interface by growing the films on substrates with evaporated Cr/Au contacts (in 2-wire configuration). The measurements on the top and bottom surfaces were compared for selected compact samples obtaining results in agreement for the two configurations. All electrical measurements have been performed on samples grown on glass. Optical transmittance and reflectance spectra were acquired with a UV-vis-nIR PerkinElmer Lambda 1050 spectrophotometer with a 150 mm diameter integrating sphere. The spectra were normalized to correct for substrate contribution by setting to 1 the intensity at the glass/film interface. To investigate the role played by oxygen in the film structure, annealing treatments in air (500$^\circ$ C, 2 hours) were performed in a Lenton muffle furnace.

\section{Results}
\subsection{Film morphology and structural properties}
AZO films grown by PLD at different oxygen pressures show a change in morphology due to the effects of the interaction of the ablated species with the background gas, as discussed elsewhere \cite{Gondoni2011}.

In particular, as the pressure is increased, a transition from compact films (with a columnar or pseudocolumnar structure) to porous films constituted by nanoparticles assembled in hierarchical, tree-like structures is observed, similarly to what we reported for other metal oxide films grown by PLD \cite{Fonzo2009}. Figure \ref{fig:sem} shows the trend in morphology for O$_2$ pressures from 0.01 Pa to 200 Pa. At low pressures SEM images show oriented columnar domains with variable size, at intermediate pressures the morphology becomes granular, and at high pressures porous, open structures with an increasing fraction of voids are clearly visible.  These differences can be explained by taking into account the collisions between the ablated species and the oxygen molecules, which allow the formation of clusters within the ablation plume and decrease the kinetic energy of the deposited species\cite{Geohegan1998}. The transition between these growth regimes is found to occur at O$_2$ pressures between 10 Pa and 100 Pa. A measurement of deposition rate with a quartz microbalance, upon comparing the thicknesses measured by SEM, revealed a decrease in film density from bulk value to less than 1 g/cm$^3$ for compact and porous samples, respectively. Film thickness was maintained around 500 nm for compact films and 2 \micro m for porous films (pressures of 100 Pa and higher). A high-resolution image of a porous sample grown at 160 Pa is reported in figure \ref{fig:sem}(g-h). An image of the same sample after annealing in air is also provided, to highlight the negligible effects of grain growth on mesoscale morphology. \\
Structural characterization via X-ray diffraction (not shown here, see ref. [\cite{Gondoni2011}]) pointed out the presence of a preferential growth direction along the $c$-axis of the ZnO wurtzitic structure for compact films. A thorough discussion of XRD patterns of AZO compact films grown at O$_2$ pressures up to 10 Pa can be found in our previous work \cite{Gondoni2011}, where the trend in vertical domain size for compact samples (calculated from Scherrer's formula on the (002) peak) shows a maximum of 30 nm at 2 Pa. At lower pressures domain size drops until reaching about 4 nm at 0.01 Pa, and it decreases also at higher pressures (10 nm at 10 Pa). A progressive shift at lower pressures in the position of the (002) peak with respect to bulk ZnO was also found, suggesting deformations in the lattice which can indicate the presence of oxygen vacancies \cite{Dong2008}.
The preferential orientation is partially lost at pressures of 10 Pa, where the most intense reflexes typical of polycrystalline ZnO appear along with the dominant (002) peak, indicating a transition towards random orientation of the nanocrystals. At higher pressures the preferential orientation is completely lost and the diffraction peaks are superimposed on a slowly decreasing background (not shown) typical of disordered structures (as a consequence of the abundance of grain boundaries).

\subsection{Electrical properties}
\begin{figure}
    \centering
        \includegraphics[width=.45\textwidth]{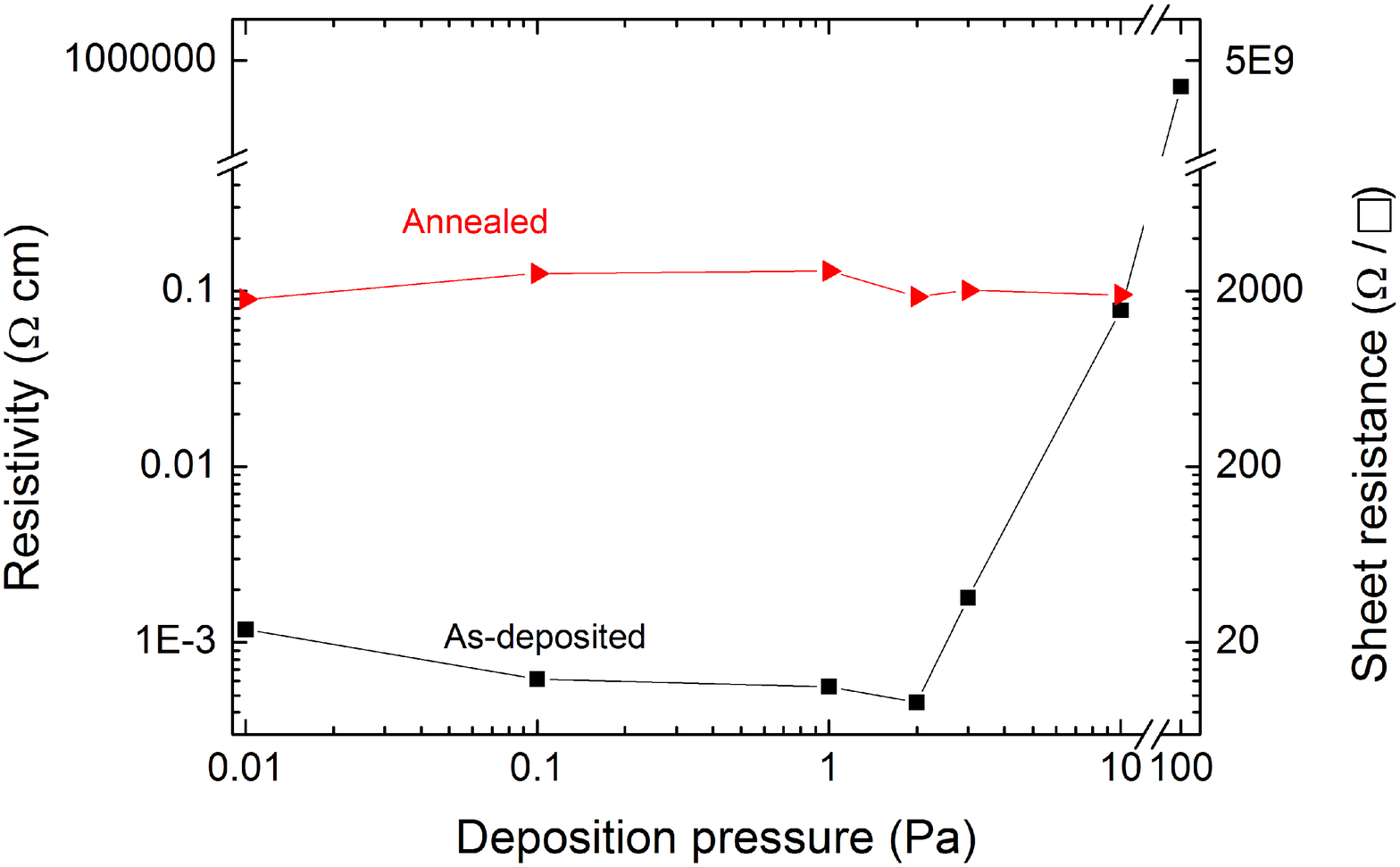}
    \caption{Film resistivity and sheet resistance as a function of deposition pressure. Black dots represent as-deposited films, red triangles are for annealed samples.}
    \label{fig:Rho}
\end{figure}
The resistivity and sheet resistance values measured for samples grown at pressures up to 100 Pa are shown in figure \ref{fig:Rho}. Compact samples are conductive, with state-of-the-art \cite{Herrero2010, Anders2010, Dong2008} resistivity values initially decreasing from $10^{-3}\,\Omega\,$cm to $4\cdot 10^{-4}\,\Omega\,$cm as the pressure is increased from 0.01 Pa to 2 Pa, and then abruptly increasing up to about 0.1 $\Omega$ cm at 10 Pa. This corresponds to sheet resistance values of the order of 10-20 $\Omega / \square$ at 2 Pa, increasing up to 2000 $\Omega / \square$ at 10 Pa. Porous samples exhibited an insulating behaviour, as the resistivity at 100 Pa is of the order of 1 M$\Omega\,$cm and it is expected to increase at higher pressures. The resistivity of compact films was measured after a thermal treatment in air to induce saturation of oxygen vacancies; an increase in resistivity up to 0.1 $\Omega$ cm is observed after annealing (figure \ref{fig:Rho}), with a trend which becomes insensitive to deposition conditions. \\
\begin{figure}
    \centering
        \includegraphics[width=0.5\textwidth]{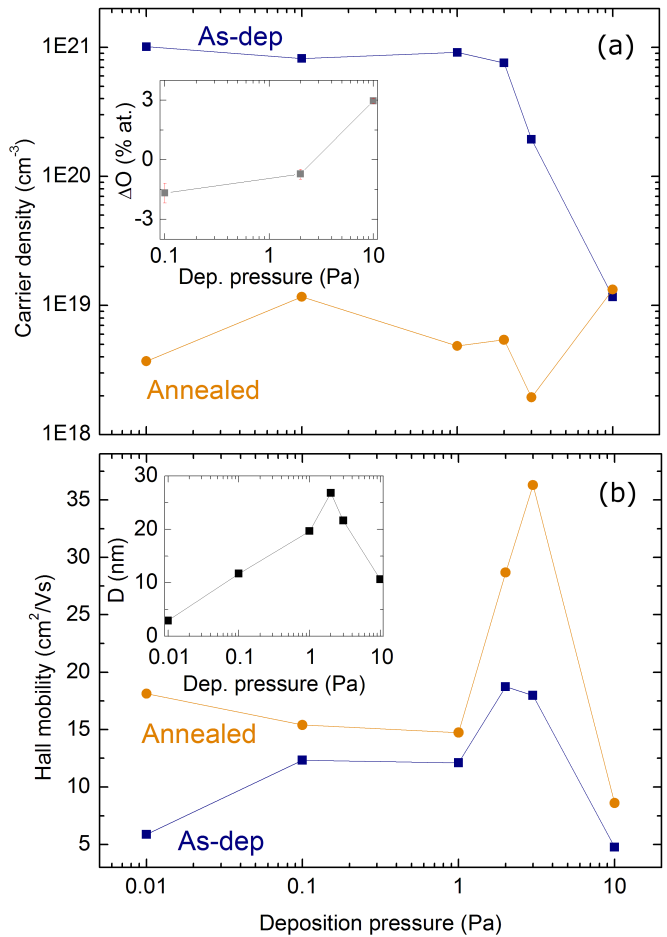}
    \caption{(a) Carrier density of compact samples grown at different O$_2$ pressures. Blue squares represent as-deposited samples, orange dots are for annealed samples. The inset shows the film oxygen content with respect to the AZO target as measured by EDXS. \\
    (b) Hall mobility of compact samples. The inset shows the vertical domain size from Scherrer's formula on the (002) XRD peak.}
    \label{fig:mun2}
\end{figure}

\begin{figure}[h]
    \centering
        \includegraphics[width=0.45\textwidth]{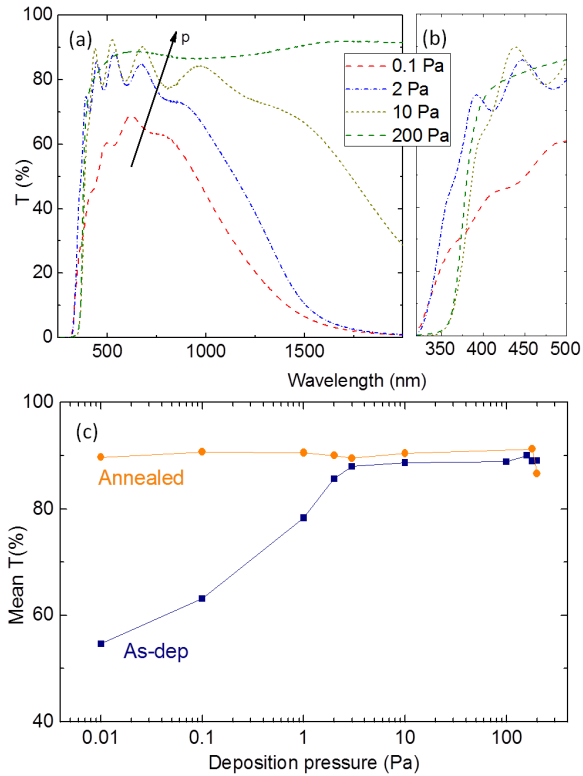}
    \caption{(a) Total transmittance spectra of as-deposited films grown at different O$_2$ pressures. (b) A close-up of the onset of absorption in the near UV - visible range. (c) Mean transmittance values in the 400 nm - 700 nm range for both as-deposited and annealed films, as a function of O$_2$ deposition pressure.}
    \label{fig:T}
\end{figure}

Hall effect measurements allowed to analyze the contribution of conduction band carrier concentration and mobility to the trends discussed above; the results are presented in figure \ref{fig:mun2}. Carrier concentration ($n$-type) decreases monotonically as the deposition pressure is increased, from $\approx10^{21}$ cm$^{-3}$ at 0.01 Pa to $\approx10^{19}$ cm$^{-3}$ at 10 Pa. Annealing in air causes the concentration of free carriers to decrease significantly, reaching a value of the order of 10$^{19}$ cm$^{-3}$ which is quite insensitive to deposition conditions: this is a further indication of the role played by oxygen vacancies, as discussed in the following sections. For this purpose, the top inset in figure \ref{fig:mun2} shows the shift in O content with respect to the deposition target (i.e. film oxygen content minus target oxygen content), as estimated from EDXS measurements. A significant increase in oxygen content is measured, indicating that films grown at low O$_2$ pressures are oxygen deficient. The increase in O content is also possibly enhanced by adsorption of oxygen from the environment due to porous morphology of films grown at 10 Pa O$_2$. \\
The trend in Hall mobility, both for as-deposited and annealed samples, presents a maximum at 2-3 Pa, where the mobility values exceed 15 cm$^2$/Vs before annealing and 30 cm$^2$/Vs after annealing. Moving away from the optimal deposition pressures the values are lower, reaching 5 cm$^2$/Vs and 7 cm$^2$/Vs before and after annealing, respectively. In general, mobility appears uniformly higher after annealing. The trend in as-deposited carrier mobility and that of mean domain size calculated from Scherrer's formula on the (002) XRD peak (bottom inset) are in agreement, suggesting the importance of defects and grain boundary scattering in limiting electrical transport.

\subsection{Optical properties}
\begin{figure}
    \centering
        \includegraphics[width=.45\textwidth]{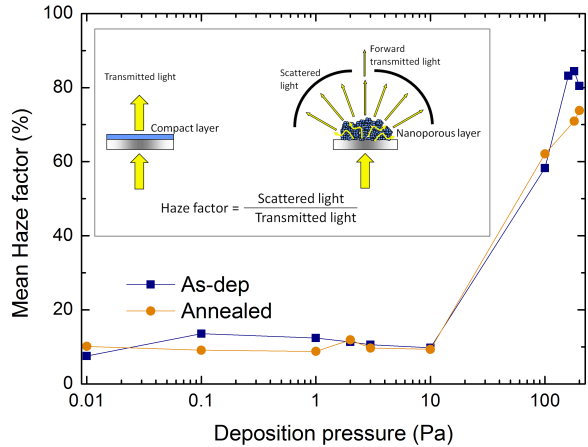}
    \caption{Mean haze factor in the visible range (400 nm - 700 nm) for as-grown (blue squares) and annealed (orange dots) samples. The inset shows a scheme of the measurement.}
    \label{fig:haze}
\end{figure}

Optical total transmittance spectra for samples grown on glass are shown in figure \ref{fig:T}, where the spectra have been corrected for the glass contribution by setting to 1 the intensity at the glass/film interface.
The spectra show an increase in transparency with increasing deposition pressure, as observed in figure \ref{fig:T}. In the region below 300 nm, where photons excite interband electronic transitions, films grown at higher pressures exhibit a sharper onset of absorption. The oscillations in the visible range are due to interference phenomena, and their position depends on film thickness and refractive index. The spectra of porous sample do not present interference fringes, due to mesoscale disorder and increased surface roughness. The near-infrared region is characterized by a decrease in optical transmittance due to carrier absorption in the conduction band for samples grown at low O$_2$ pressures (up to 10 Pa) while porous films show high transmittance up to 2000 nm. \\
A comparison of the mean values of optical transmittance in the visible range, taken between 400 nm and 700 nm, is reported in figure \ref{fig:T}(c) for as-deposited and annealed samples. For as-deposited films (squares in fig.\ref{fig:T}(c)) the values increase monotonically with pressure, from 55\% at 0.01 Pa up to nearly 90\% at 10 Pa. Porous films are characterized by an overall higher transparency, in that the total transmittance is comparable but the thickness is 4 times as much.
Transmittance values for annealed films (dots in fig.\ref{fig:T}(c)) are sensibly higher than prior to the treatment for compact samples. The transmittance of porous samples does not vary significantly upon annealing. In general, the variation of transmittance versus deposition pressure for annealed samples is almost negligible, as was the case for electrical properties.
\\
Light scattering properties were characterized by means of haze factor measurements, i.e. the ratio of diffuse transmittance to total transmittance (see inset in fig. \ref{fig:haze}). Diffuse transmittance was measured by letting the unscattered light fraction out of the integrating sphere through a slit.
Average values in the 400 nm - 700 nm range are reported in figure \ref{fig:haze}. The haze factor of compact films is around 10\%, for deposition pressures up to 10 Pa. Above 10 Pa it increases reaching values of 80\% and beyond: porous films show improved light scattering properties, revealing that over 80\% of the transmitted photons experience scattering phenomena. We remark that mean transmittance of porous samples in the same range is over 80\%. The effects of annealing on haze factor are almost negligible, as morphology at the mesoscale is unaffected by such treatment (see fig. \ref{fig:sem}).

\section{Discussion}
Films grown at low oxygen pressure (0.01 Pa - 3 Pa) have a compact structure characterized by a high concentration of oxygen vacancies. This has an effect on electrical properties: each oxygen vacancy acts as a doubly charged electron donor, which contributes to free carrier density (fig. \ref{fig:mun2}), and its contribution is lost with vacancy saturation upon annealing. As for optical properties, the absorption due to interband transitions, (fig. \ref{fig:T}(b)) in the 300 nm - 500 nm range, is characterized by two different regimes: compact, oxygen-deficient films are more transparent in the higher-energy region (up to 380 nm) and porous, oxygen-rich samples are more transparent in the blue. This behaviour is in excellent agreement with results from \textit{ab initio} calculations: for example, K\"{o}rner \textit{et al.}\cite{Korner2010} have shown that grain boundaries create O $2p$ dangling bonds, i.e. available states above the valence band maximum which decrease the optical gap. O vacancies allow for saturation of these dangling bonds, causing the optical gap to reopen and increasing UV transmittance of films grown at low O$_2$ pressures. In the visible, O vacancies increase light absorption through intragap states. In the IR region, absorption due to plasma oscillations of free electrons in the conduction band (from shallow states below conduction band minimum) is extremely strong for samples grown at low pressure which have higher concentrations of electrons in the conduction band. \\
\begin{figure}[h!]
    \centering
        \includegraphics[width=0.45\textwidth]{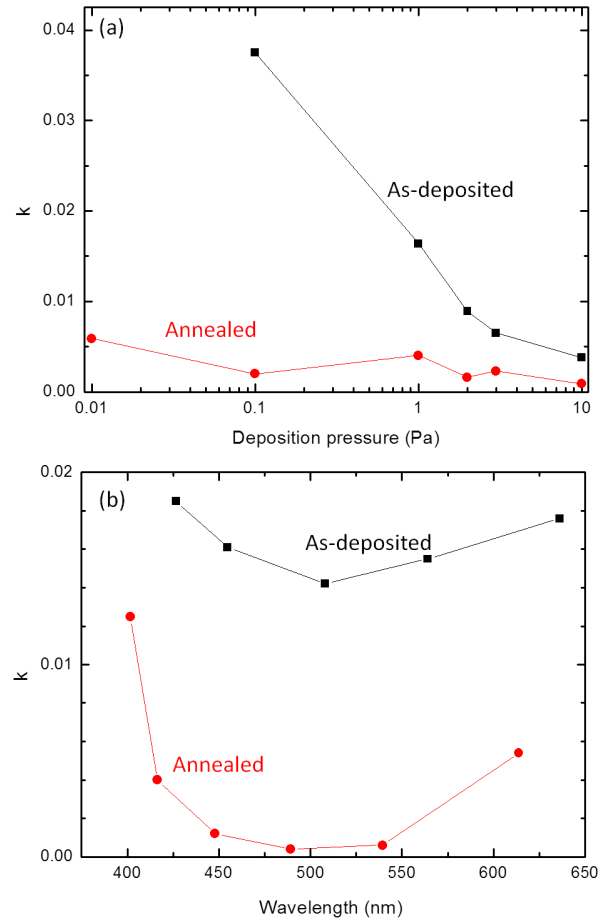}
    \caption{(a) Extinction coefficient of a sample grown at 1 Pa as a function of incoming wavelength. The curve is show before (squares) and after (dots) annealing. (b) Mean extinction coefficient of as-deposited (squares) and annealed (dots) samples, as a function of deposition pressure}
    \label{fig:k}
\end{figure}
As the oxygen pressure is increased (10 Pa - 100 Pa), local stoichiometry order is achieved: O vacancies do not contribute to free carriers at 10 Pa, which results in higher resistivity and improved transparency due to the disappearing of below bandgap optical transitions and the weakening of plasma absorption. This is also confirmed by EDXS measurements, according to which from 10 Pa the oxygen content within the films equals that of the deposition target. At the same time, the morphology of such films is evolving towards granular-slightly porous: this is detrimental to domain size which is limited by the size of the building blocks constituting the hierarchical structure (as found from XRD measurements) and hence to carrier mobility (see fig. \ref{fig:mun2}).\\
The poor conductivity of porous films (100 Pa - 200 Pa) can be explained by taking into account the mesoporous morphology visible in fig. \ref{fig:sem} and its effects on charge carrier transport in the in-plane direction, which may result in different current carrying behaviour (e.g., hopping). It may be reasonable to expect better electrical transport properties in the cross-plane direction, but this aspect constitutes an open problem due to the difficulty of applying electrical contacts on the top of delicate porous films. Optical transparency is high throughout the spectra, as an evidence of a less defective local structure, and a significant presence of voids. \\
The effects of annealing in air allow to study the contribution of O vacancies to conduction electrons: the drop in carrier concentration in fig. \ref{fig:mun2} and the increase in IR transmittance in fig. \ref{fig:T}(a) are in agreement with one another. The slight but uniform increase in mobility (fig. \ref{fig:mun2}) is a possible consequence of improved structural order upon annealing. \\
The overall behaviour of electrical properties indicates that at low deposition pressure, O vacancies significantly contribute to conduction electron density, whose mobility is dictated by morphology and domain size at higher pressures. In this respect, several authors \cite{Makino2006,Minami2005} have pointed out how the adsorption of oxygen atoms at grain boundaries in AZO can cause the trapping of charge carriers and further decrease their mobility.
At any rate, it is reasonable to expect significant interplay between dopants and oxygen vacancies as far as concerns electrical properties; we are currently unable to investigate the position of Al$^{3+}$ ions and O vacancies in the lattice, and its effect on carrier concentration and mobility.\\
The method of envelope analysis \cite{Magnifacier1976,Swanepoel1984} was employed to estimate the refractive index of compact films by numerically solving the envelope equations for the position of the interference fringes in transmittance spectra, knowing film thickness as measured by SEM. The results of our calculations indicate normal dispersion relations for $n$ (not shown), and the mean values of the real part of the index in the visible range is found to be about 1.90 for all compact films (pressures up to 10 Pa), which is in agreement with the typical values for ZnO (1.8-2.4)\cite{Ozgur2005}. The variation of the extinction coefficient was also estimated over the visible spectrum (fig. \ref{fig:k}(a)) and its mean value in the 400 nm - 700 nm range as a function of deposition pressure is reported in fig. \ref{fig:k}(b).
The calculated dispersion relations of the imaginary part $k$ show a minimum in the visible and a dependence on deposition pressure following the trends found in transmittance: films grown at higher oxygen pressures are characterized by a higher degree of structural order resulting in less significant light absorption. The extinction coefficient of samples grown at low pressures decreases by a factor 10 upon annealing, giving confirmation of the effect of defect healing on transparency. It is also worthwile to notice that there is qualitative agreement between this estimate of $k$ and an estimate of the absorption coefficient $\alpha$ taken from Lambert-Beer's law, as the relation $\alpha = 4\pi k /\lambda$ holds (substituting the average values yields $\alpha \approx 10^4$ cm$^{-1}$). \\

\begin{figure}[h]
    \centering
        \includegraphics[width=.45\textwidth]{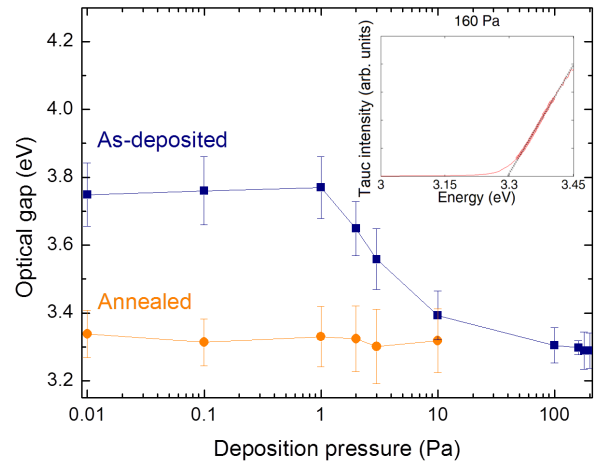}
    \caption{Optical gap vs. deposition pressure. The inset shows an example of Tauc plot and linear fit.}
    \label{fig:gap}
\end{figure}

The optical gap was measured by means of Tauc plots
\cite{Tauc1968}, by plotting $(\alpha h \nu)^2$ vs. $h\nu$ in
proximity of the onset of absorption, taking $\alpha$ from
transmittance spectra via Lambert-Beer's law (the relation $\alpha
= -\ln T / d$, where $d$ denotes film thickness, was used since
reflectance was negligible in this spectral range) and performing
a linear fit in the region where $(\alpha h \nu)^2 \propto
\sqrt{h\nu - E_g}$, where $E_g$ is the optical gap. The intercept
of the linear fit thus gives $E_g$. The choice of using $(\alpha
h\nu)^2$ in the plots is due to the direct bandgap of ZnO and to
the estimated filling of the conduction band, as discussed by
Buchholz \textsl{et al.}\cite{Buchholz2009}. The behaviour of
$E_g$ as a function of deposition pressure is reported in figure
\ref{fig:gap}, where the inset shows an example of Tauc plot for a
porous sample.

\begin{figure}[h]
    \centering
        \includegraphics[width=.45\textwidth]{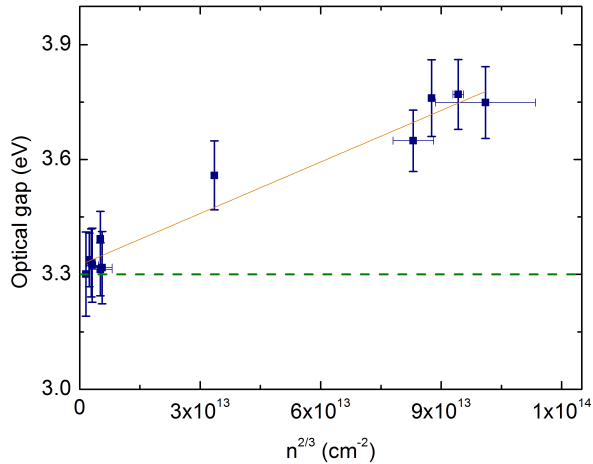}
    \caption{Plot of $n^{\frac{2}{3}}$ vs. $E_g$ and linear interpolation. The dashed line shows the optical gap of intrinsic ZnO.} \label{fig:mbu}
\end{figure}

\begin{figure}[h]
    \centering
        \includegraphics[width=.45\textwidth]{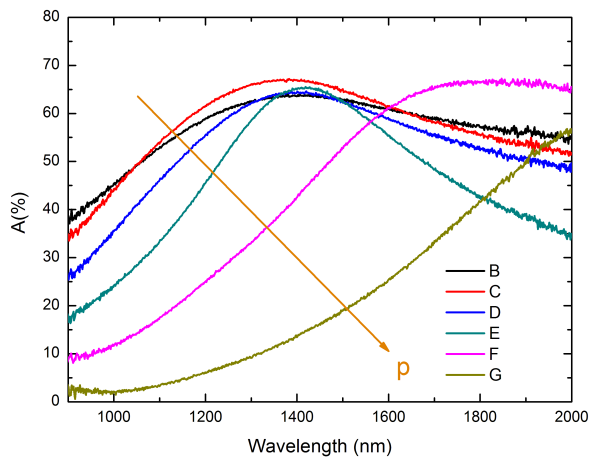}
    \caption{Absorptance (\%) of compact films in the plasma region for different growth pressures.}
    \label{fig:assir}
\end{figure}

The optical gap of as-deposited films shows a decreasing trend from 3.75 eV to 3.3 eV (which equals the optical gap of bulk undoped ZnO\cite{Ozgur2005}) as the O$_2$ deposition pressure is increased, whereas annealed samples have a smaller optical gap, with only weak dependence on deposition pressure. This is an evidence of Moss-Burstein effect (MB) \cite{Burstein1954}. As the conduction band is filled with electrons (which we know from electrical measurements), the optical gap increases proportionally to $n^{\frac{2}{3}}$, where $n$ is the conduction band electron density. Figures \ref{fig:gap} and \ref{fig:mun2} indicate that the decrease in optical gap follows the decrease in carrier concentration. Since the shift in optical gap obeys the equation
\begin{equation}
    \Delta E_g=\frac{\hbar^2}{2m^*}\left(3\pi^2n\right)^\frac{2}{3}
\end{equation}
a plot of $E_g$ vs. $n^{\frac{2}{3}}$ can be used to estimate the electron effective mass $m^*$, provided that the extrapolation for $n=0$ gives the optical gap for intrinsic ZnO, i.e. 3.3 eV. The plot is shown in figure \ref{fig:mbu}, in which the linear interpolation has an intercept given by $E_{g,0} = 3.33$ eV. The agreement of this value with the literature is an indication of the reliability of the fit, nonetheless the degree of uncertainty due to the nonuniform distribution of experimental data remarks that the result is only an estimate.
From the slope of the linear interpolation it is possible to estimate an electron effective mass of $m^*=0.81\,m_0$, where $m_0$ is the electron rest mass. This value is likely to overestimate the actual effective mass because of many-body effects which tend to narrow the bandgap in opposition to the MB effect \cite{Sernelius1988, Jain2006}. Other estimates of $m^*$ in AZO calculated from MB shift are even higher ($m^*\approx 0.98m_0$)\cite{Dong2008} which points out other fine structure effects possibly due to nonparabolicity of the bands. \\
The plasma infrared absorption peaks of compact samples were used to obtain another estimate of the electron effective mass. Absorption profiles were taken from transmittance and reflectance spectra (by taking $1-T-R$) and are reported in figure \ref{fig:assir}.
The plasma wavelength (i.e. the wavelength at which photons are absorbed to sustain collective oscillations of free electrons in the conduction band) is seen to increase as the deposition pressure is increased, and the peak width is smallest at 2 Pa: it has already been reported in literature \cite{Coutts2000} that increasing the conductivity of a TCO film results in a narrower plasma absorption in the IR region.
Since the position of the peak is given by \cite{Coutts2000}
\begin{equation}\label{eq:index}
        \omega_p=\sqrt{\frac{ne^2}{m^*\varepsilon_{0}\varepsilon_{\infty}}}
\end{equation}
the redshift of the plasma frequency is readily explained with the decrease in carrier concentration which was measured by electrical measurements. By approximating the high-frequency permittivity $\varepsilon_{\infty}$ with the squared refractive index, eq. \ref{eq:index} was used to estimate the electrons conductive band effective mass, finding $m^*\approx 0.4 m_0$. We remark that this value is close to other estimates of $m^*$ in AZO calculated from plasma absorption ($m^*\approx 0.5-0.6\,m_0$) \cite{Qiao2006, Brehme1999} and that no author so far has provided two independent estimates of $m^*$ to the best of our knowledge.\\

The discrepancy between this value and the one obtained from MB effect measurements can be due to several factors: plasma oscillations mainly interest the bottom of the conduction band whereas optical transitions reach points farther from the center of the Brillouin zone; second, the value calculated from Moss-Burstein shift takes into account annealed films which do not absorb in the IR and thus includes a range of materials with a different structure, furthermore the approximated value of $\varepsilon_{\infty}$ can introduce some uncertainty.

\section{Conclusions}
We have provided a study of structural, electrical and optical properties of nanostructured Al-doped ZnO thin films. By pulsed laser ablation at different O$_2$ pressures a variety of structures were grown, from compact transparent conductors (0.01 - 10 Pa) to nanoparticle assemblies with a hierarchical forest-like structure (100 - 200 Pa). We showed that compact films are highly performing TCOs with low resistivity (4.5$\cdot 10^{-4}\,\Omega$cm) and high transparency ($>80\%$ in the visible range) and discussed how this is due to a compromise between the presence of oxygen vacancies and that of mobility-limiting defects. Porous films grown at higher pressures exhibit prominent transparency and light scattering properties (both transmittance and haze over 85\% in the visible range) but show poorer conductivity ($\approx 1$\micro S/cm) due to the open structure with distributed connectivity. \\
By combining the results of Hall effect, EDXS and XRD measurements we were able to demonstrate how the trend in conduction electron density of compact films is influenced by O concentration, whereas their mobility is limited by grain boundary scattering, hence its trend follows domain size. The in-plane insulating behaviour of porous sample is ascribed to morphology effects, i.e. the lack of continuous paths for charge carriers.
The effects of annealing in air were studied to investigate the role of O vacancy saturation and defect healing: the dependence on deposition parameters was nearly lost for all compact films, which suffered from a significant decrease in conductivity (up to 2 orders of magnitude) upon saturation of oxygen vacancies. Porous samples were basically unaffected by annealing, as their functional properties are mainly morphology-driven, and morphology was left unchanged by the thermal treatment.
By investigating the Moss-Burstein shift exhibited by the films we were able to provide an estimate of the electron conduction band effective mass ($m^* \approx 0.81\,m_0$). An indipendent estimate derived from carrier infrared absorption and index of refraction was also obtained, finding $m^*\approx 0.4\,m_0$: we discussed the possible (physical and experimental) reasons for this discrepancy.\\
We believe that a detailed understanding of the
structure-dependent functional properties may open the way to
non-conventional employment of AZO in actual energy conversion
devices. An intelligent management of the incident light driven by
a fine tuning of electrodes morphology at the nano- and mesoscale
is of potential interest to improve performance of new generation
solar cells.

\end{document}